\documentclass{PoS}

\title{Evolution of minimum-bias parton fragmentation in nuclear collisions}

\ShortTitle{Evolution of parton fragmentation in nuclear collisions}

\author{\speaker{Thomas A. Trainor}\\
        CENPA 354290, University of Washington, Seattle, WA  98195\\
      E-mail: \email{trainor@hausdorf.npl.washington.edu}}

\abstract{
Hard components of $p_t$ spectra can be identified with minimum-bias parton fragmentation in nuclear collisions. Minimum-bias fragment distributions (FDs) can be calculated by folding a power-law parton energy spectrum with parametrized fragmentation functions (FFs) derived from $e^+$-$e^-$ and p-\=p collisions. Alterations to FFs due to parton ``energy loss'' or ``medium modification'' in Au-Au collisions are modeled by adjusting FF parametrizations consistent with rescaling QCD splitting functions. The parton spectrum is constrained by comparison with a p-p $p_t$ spectrum hard component. The reference for all nuclear collisions is the FD derived from in-vacuum $e^+$-$e^-$ FFs. Relative to that reference the hard component for p-p and peripheral Au-Au collisions is found to be {\em strongly suppressed} for smaller fragment momenta. At a specific point on centrality the Au-Au hard component transitions to enhancement at smaller momenta and suppression at larger momenta, consistent with FDs derived from  medium-modified $e^+$-$e^-$ FFs.
}

\FullConference{High-pT Physics at LHC -09\\
		 February 4-7 2009\\
		 Prague, Czech Republic}

\def\bea{\begin{eqnarray}}
\def\eea{\end{eqnarray}}
\def\text{}

\begin{document}

\section{Introduction}

RHIC collisions are commonly described in terms of two themes: hydrodynamic (hydro) evolution of a thermalized bulk medium and energy loss of energetic partons (hard probes) in that medium. Hydro is thought to dominate $p_t$ spectra below 2 GeV/c, parton fragmentation is expected above 5 GeV/c, and ``quark coalescence'' is thought to dominate the intermediate $p_t$ interval.

Recent studies of spectrum and correlation structure have revealed interesting new aspects of RHIC collisions. Number and $p_t$ angular correlations in the final state contain {\em minijet} structures (minimum-bias parton fragmentation)~\cite{ppcorr1,ppcorr2,axialci,hijscale,ptscale,ptedep,daugherity}. Two-component analysis of p-p and Au-Au spectra reveals a corresponding {\em hard component} interpreted as a minimum-bias fragment distribution, suggesting that jet phenomena extend down to 0.1 GeV/c hadron momentum~\cite{ppprd,hardspec}.

Minijets (well described in p-p collisions by PYTHIA/HIJING~\cite{minijet}) are observed to dominate the transverse dynamics of nuclear collisions at energies above $\sqrt{s_{NN}} \sim$ 15 GeV. The term ``minijets'' can be applied collectively to hadron fragments from the minimum-bias scattered-parton spectrum  averaged over a given A-A or N-N event ensemble. Minijets provide {unbiased} access to fragment distribution structure down to a small cutoff energy for scattered partons (those partons fragmenting to charged hadrons) and to the smallest detectable fragment momenta ($\sim 0.1$ GeV/c). 

In this analysis minijets are studied in the form of $p_t$-spectrum hard components isolated via the two-component spectrum model. Measured hard components are compared with calculated fragment distributions obtained by folding parton spectra with fragmentation-function ensembles. Parton spectrum parameters and modifications to fragmentation functions in more-central Au-Au collisions are inferred~\cite{evolve}. The goal is a comprehensive QCD description of all nuclear collisions.

\section{Two-component spectrum model}

The two-component model of p-p spectra~\cite{ppprd} is the starting point for the fragmentation analysis described here. The two-component (soft+hard) model was first obtained from a Taylor-series expansion on {\em observed} event multiplicity $\hat n_{ch}$ ($\leq$ corrected $n_{ch}$) of spectra for several multiplicity classes. The soft component was subsequently interpreted as longitudinal projectile-nucleon fragmentation, the hard component as transverse scattered-parton fragmentation. The two-component model applies to two-particle correlations on $(y_t,y_t)$ as well as their 1D projections onto $p_t$ or $y_t$. 


The two-component spectrum model for p-p collisions with corrected soft and hard multiplicities $n_s + n_h = n_{ch}$ is
\bea \label{twoc}
 \frac{1}{n_s(\hat n_{ch})}\frac{1}{y_t}\, \frac{dn_{ch}(\hat n_{ch})}{dy_t } =  S_0(y_t)  +  \frac{n_h(\hat n_{ch})}{n_s(\hat n_{ch})}\,  H_{0}(y_t),
\eea
where soft component $S_0(y_t)$ is the Taylor series ``constant,'' and hard component $H_0(y_t)$ is the coefficient of the term linear in $\hat n_{ch}$, both normalized to unit integral. $S_0(y_t)$ is a L\'evy distribution on $m_t$, $H_0(y_t)$ is a Gaussian plus QCD power-law tail on transverse rapidity $y_t = \ln\{(m_t + p_t) / m_0\}$. To compare with A-A spectra we define $S_{pp} =(1/y_t)\, dn_s/dy_t$ with reference model $n_s\, S_0$ and similarly for $H_{pp} \leftrightarrow n_h\, H_0$. The two-term Taylor series exhausts all significant p-p spectrum structure.

\begin{figure}[h]
\includegraphics[width=.24\textwidth,height=.244\textwidth]{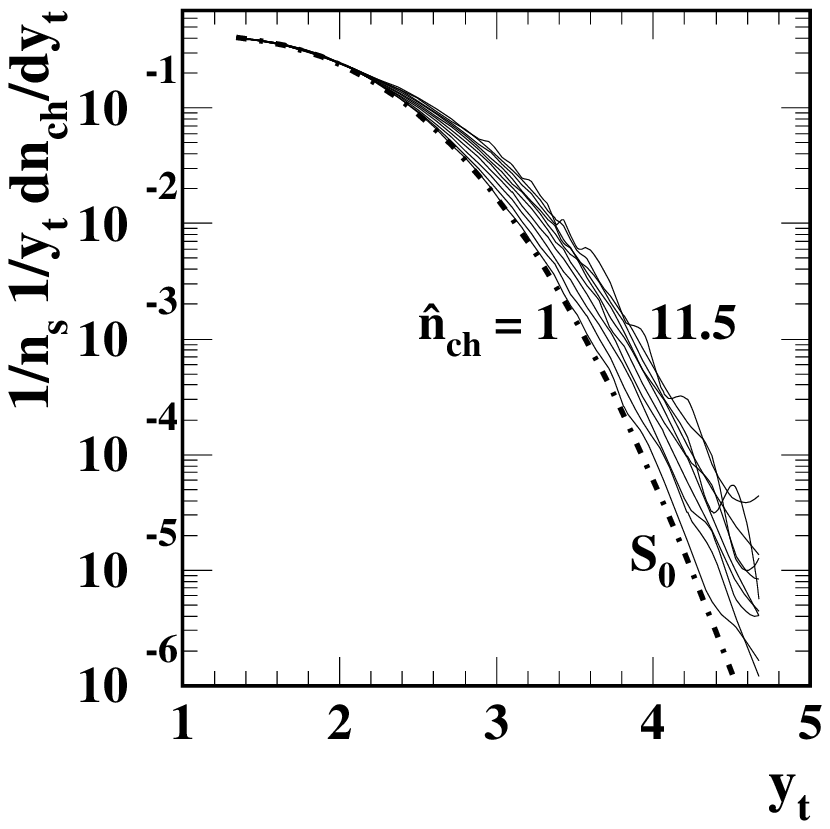}
\includegraphics[width=.24\textwidth,height=.244\textwidth]{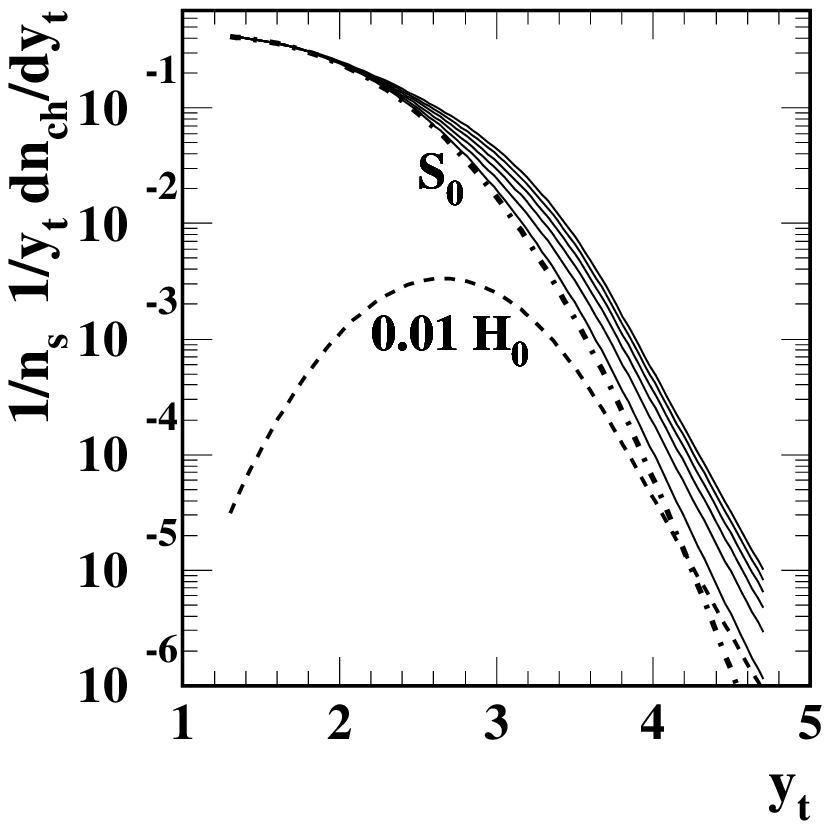}
\includegraphics[width=.48\textwidth,height=.24\textwidth]{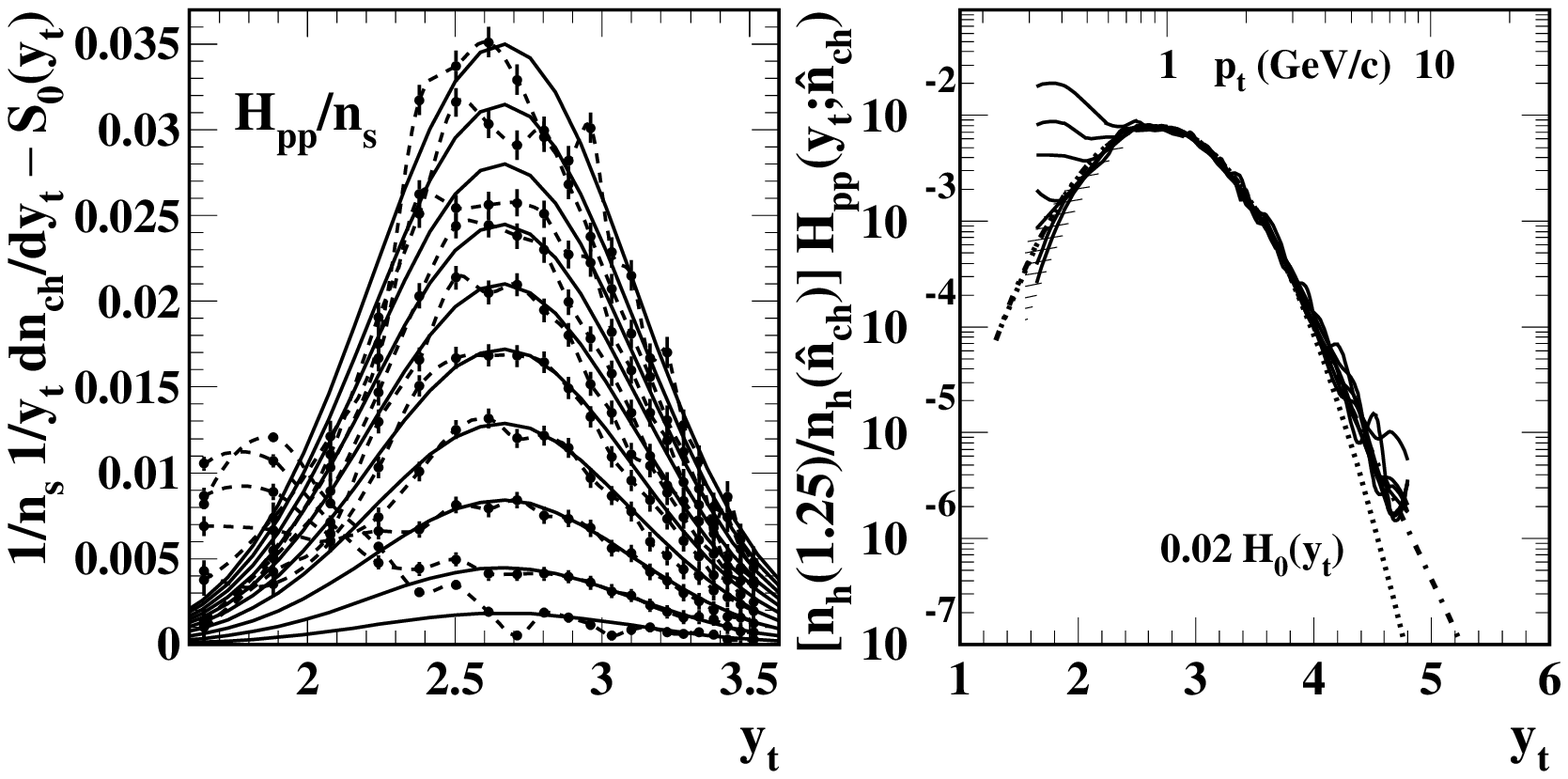}
\caption{\label{ppspec} 
First: $y_t$ spectra from $\sqrt{s_{NN}} = 200$ GeV p-p collisions for ten multiplicities,
Second: Corresponding two-component model,
Third: Corresponding hard components,
Fourth: Hard components normalized to NSD p-p collisions.
} 
\end{figure}

Fig.~\ref{ppspec} (first panel) shows spectra for ten multiplicity classes from 200 GeV  non-single diffractive (NSD) p-p collisions~\cite{ppprd}. The asymptotic limit for $\hat n_{ch} \rightarrow 0$ (dash-dotted curve) is $S_0$. The spectra are normalized by soft-component multiplicity $n_s$. Fig.~\ref{ppspec} (second panel) shows the two-component algebraic model Eq.~(\ref{twoc}) with unit-normal model functions $S_0$ and $H_0$ defined in~\cite{ppprd,hardspec}. Hard-component coefficient $n_h  / n_s$ scales as $\alpha\, \hat n_{ch}$. Factor $\alpha = 0.01$ is the average value for most $\hat n_{ch}$ classes. The spectrum data in the first panel are described to the statistical limits.

Figure~\ref{ppspec} (third panel) shows hard components $ H_{pp} / n_s$ for ten multiplicity classes obtained by subtracting fixed soft component $S_0$ from the ten NSD p-p spectra normalized to $n_s$. The shape is Gaussian independent of multiplicity~\cite{ppprd}.
Fig.~\ref{ppspec} (fourth panel) shows hard components $H_{pp}$ from the third panel scaled by factors $n_h(1.25) / n_h(\hat n_{ch})$ to obtain the mean hard component for NSD p-p collisions. The dash-dotted curve is $0.02\, H_0$ [$0.02 \sim (\alpha = 0.007)\, (\hat n_{ch}=1.25)\, (n_{s} = 2.5)$~\cite{ppprd}]. The exponential tail represents the QCD power law $\propto p_t^{-n_{QCD}}$. The spectrum hard component is interpreted as a {\em minimum-bias} fragment distribution dominated by ``minijets''---jets from those partons (gluons) with at least the minimum energy required to produce charge-neutral combinations of charged hadrons. Equivalent structure appears in two-particle correlations on $(y_t,y_t)$~\cite{ppcorr1,ppcorr2}.


The corresponding two-component model for per-participant-pair A-A spectra is 
\bea  \label{aa2comp}
\frac{2}{n_{part}} \frac{1}{y_t}\frac{dn_{ch}}{dy_t} &=& S_{NN}(y_t) +  \nu\, H_{AA}(y_t;\nu) \\ \nonumber
&=&  S_{NN}(y_t) +  \nu\,r_{AA}(y_t;\nu) \,H_{NN}(y_t),
\eea
where $S_{NN}$ ($\sim S_{pp}$) is the soft component and $H_{AA}$ is the A-A hard component (with reference $H_{NN}\sim H_{pp}$) integrating respectively to multiplicities $n_s$ and $n_h$ in one unit of pseudorapidity $\eta$~\cite{ppprd,hardspec}. Ratio $r_{AA} = H_{AA} / H_{NN}$ is an alternative to nuclear modification factor $R_{AA}$. Centrality measure $\nu \equiv 2 n_{binary} / n_{participant}$ estimates the Glauber-model mean nucleon path length. We are interested in the evolution of hard component (fragment distribution) $H_{AA}$ or ratio $r_{AA}$ with A-A centrality.

\section{Fragmentation functions}

$e^+$-$e^-$ (e-e) fragmentation functions (FFs) have been parametrized accurately over the full kinematic region relevant to nuclear collisions. e-e light-quark and gluon fragmentation functions\, $ D_{xx}(x,Q^2) \leftrightarrow D_{xx}(y,y_{max}) $ ($xx$ is the FF context: e-e, p-p, A-A) are accurately described above energy scale (dijet energy) $Q \sim 10$ GeV by a two-parameter {\em beta distribution} $\beta(u;p,q)$ on normalized rapidity $u$~\cite{ffprd}. Fragment rapidity for unidentified hadrons is $y = \ln[(E+p)/m_\pi]$, and parton rapidity $y_{max} = \ln(Q/m_\pi)$. Parameters $(p,q)$ vary slowly and linearly with $y_{max}$ above 10 GeV and can be extrapolated down to $Q \sim 4$ GeV based on dijet multiplicity data. 

\begin{figure}[h]
\includegraphics[width=.24\textwidth,height=.244\textwidth]{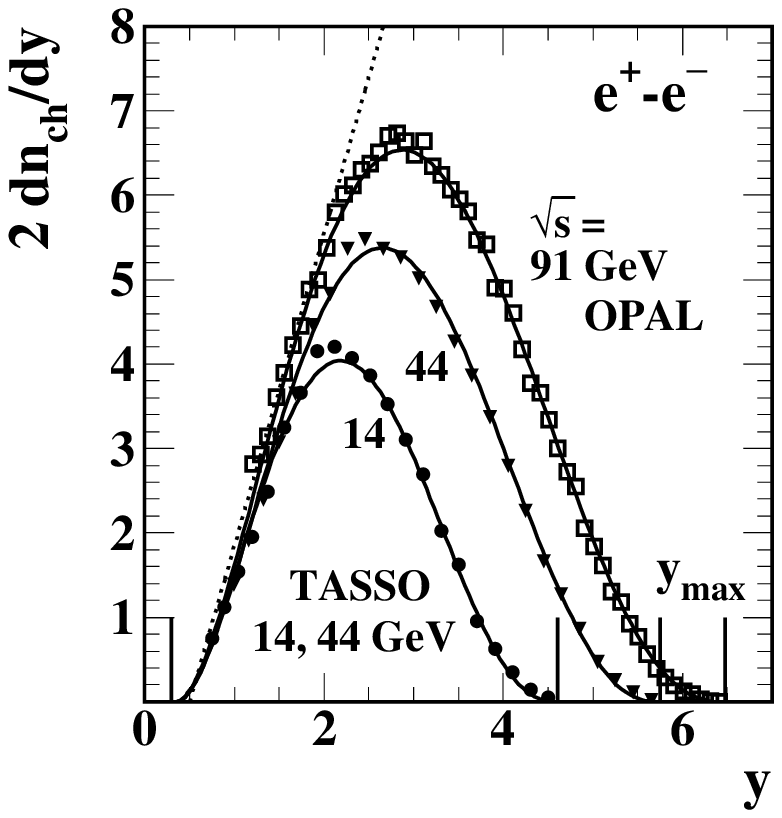}
\includegraphics[width=.24\textwidth,height=.24\textwidth]{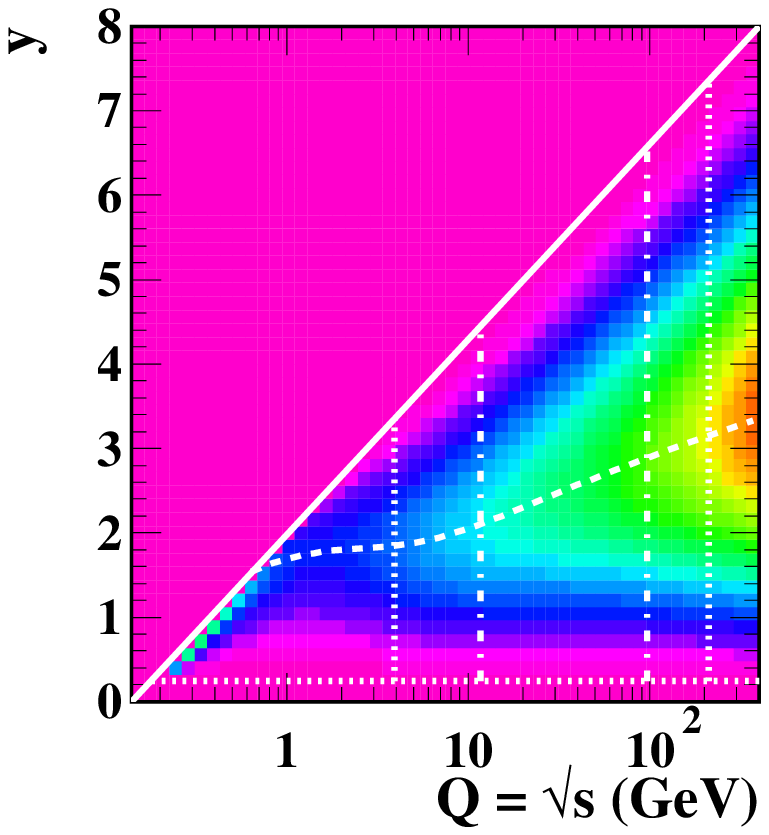}
\includegraphics[width=.24\textwidth,height=.244\textwidth]{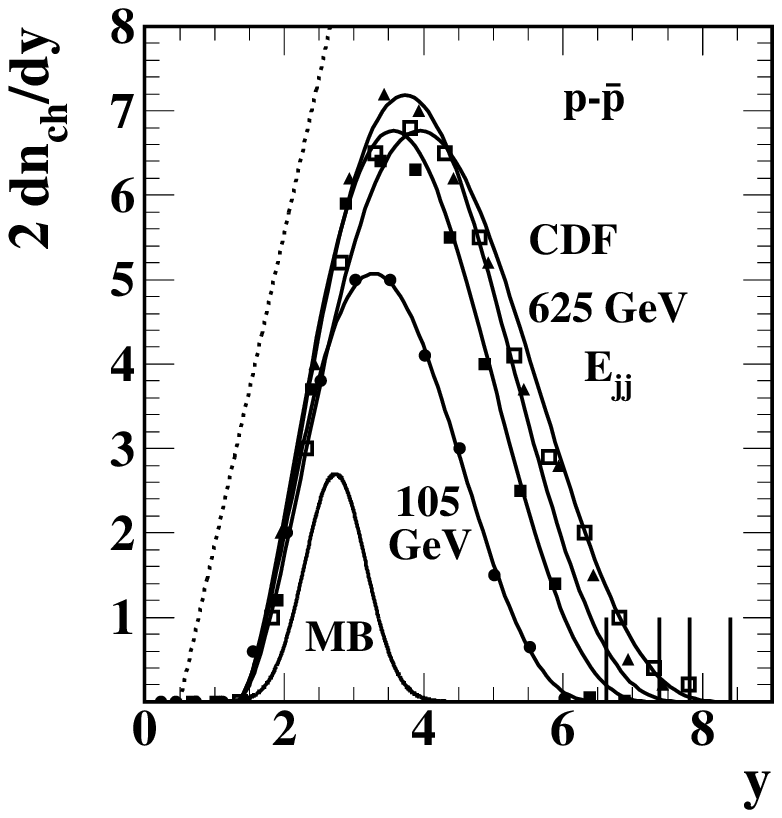}
\includegraphics[width=.24\textwidth,height=.24\textwidth]{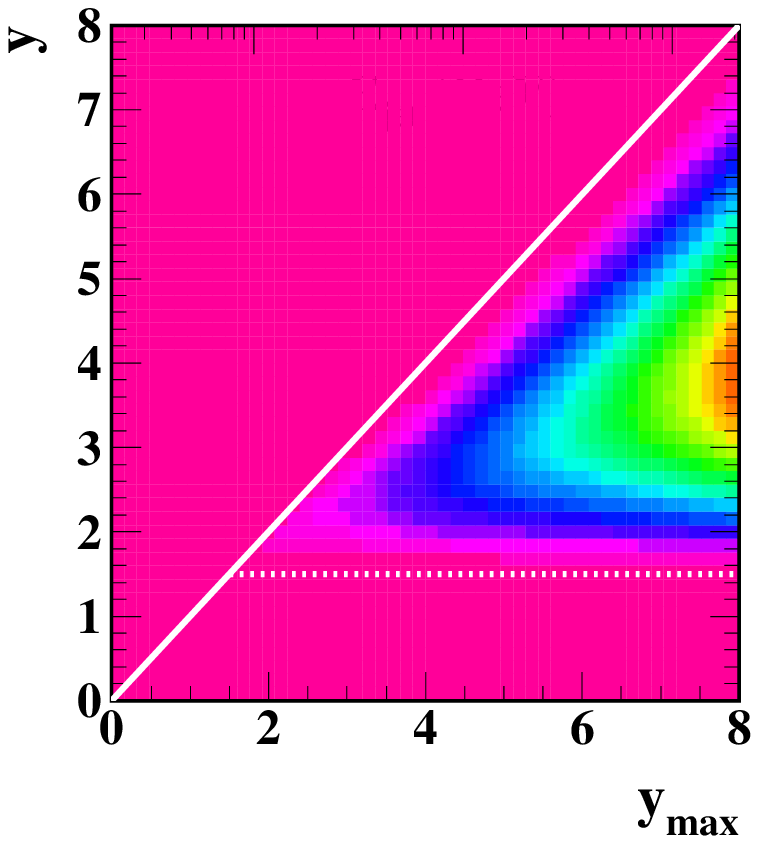}
\caption{\label{eefda} 
First: Fragmentation functions (FFs) from $e^+$-$e^-$ collisions for three energies with $\beta$-distribution parametrizations (solid curves),
Second: Full $e^+$-$e^-$ FF parametrization on parton rapidity $y_{max}$,
Third: FFs from p-\=p collisions for several dijet energies,
Fourth: Full p-\=p FF parameterization on parton rapidity.
} 
\end{figure}

Fig.~\ref{eefda} (first panel) shows measured FFs for three energy scales from HERA/LEP~\cite{tasso,opal}. The 2 in the axis label indicates {\em dijet} $n_{ch}$ densities. The vertical lines at right denote $y_{max}$ values. The curves are determined by the $\beta(p,q)$ parametrization with $y_{min} \sim 0.35$ ($p_t \sim 0.05$ GeV/c, left vertical line) and describe data to their error limits over the entire fragment momentum range. 
Fig.~\ref{eefda} (second panel) shows the FF ensemble (inclusive light quarks fragment to inclusive hadrons) vs energy scale $Q$ as a surface plot~\cite{ffprd}. The dashed curve is the {\em locus of modes}---the maximum points of the FFs. Between the dash-dotted lines the system is determined by FF data.  Between the dash-dotted and dotted lines the parametrization is constrained only by dijet multiplicities. 

Figure~\ref{eefda} (third panel) shows FF data from p-\=p collisions at FNAL (samples from the full data set)~\cite{cdf1}. The solid curves guide the eye. There is a significant systematic difference between p-p and e-e FFs. The dotted line represents the lower limit for e-e FFs. The systematic gap for all parton energies is apparent---$y_{min}$ for p-p collisions is $\sim 1.5$ (0.3 GeV/c) instead of 0.35 (0.05 GeV/c). The CDF FFs also reveal a systematic amplitude saturation or suppression at larger parton energies compared to LEP systematics. The curve labeled MB is the hard-component reference from NSD p-p collisions~\cite{ppprd}.
Fig.~\ref{eefda} (fourth panel) shows a surface plot of the p-p FF ensemble~\cite{evolve}. The surface represents the e-e FF parametrization modified by introducing cutoff factor
\bea
g_{cut}(y) = \tanh\{ (y - y_0)/\xi_y\}~~~y > y_0,
\eea
with $y_0 \sim \xi_y \sim 1.5$ determined by the CDF FF data~\cite{cdf1}. The modified FFs have not been rescaled to recover the initial e-e parton energy. The cutoff function thus represents real fragment and energy loss from p-p relative to e-e FFs. The difference implies that FFs are not universal.

\begin{figure}[h]
\includegraphics[width=.48\textwidth,height=.244\textwidth]{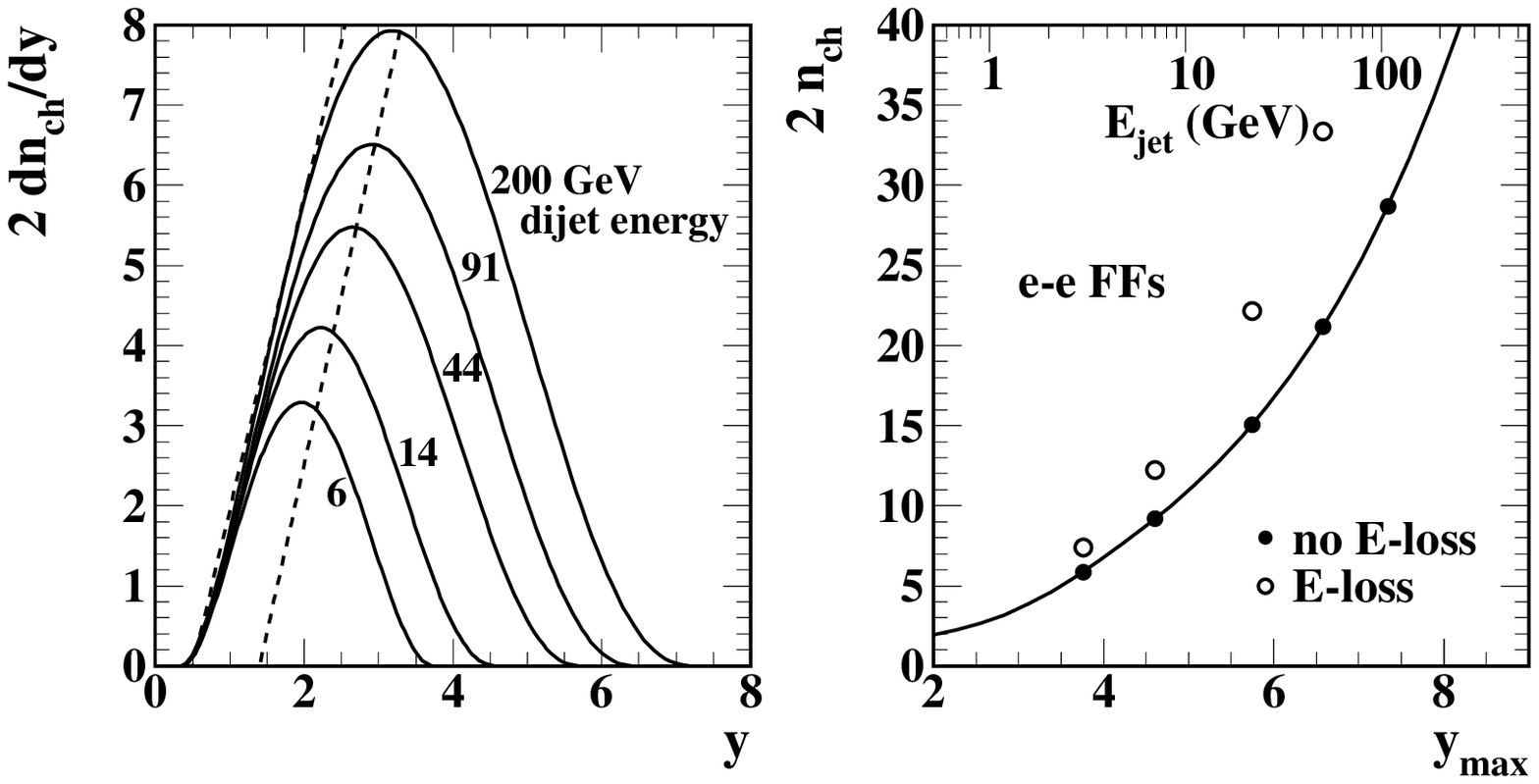}
\includegraphics[width=.48\textwidth,height=.24\textwidth]{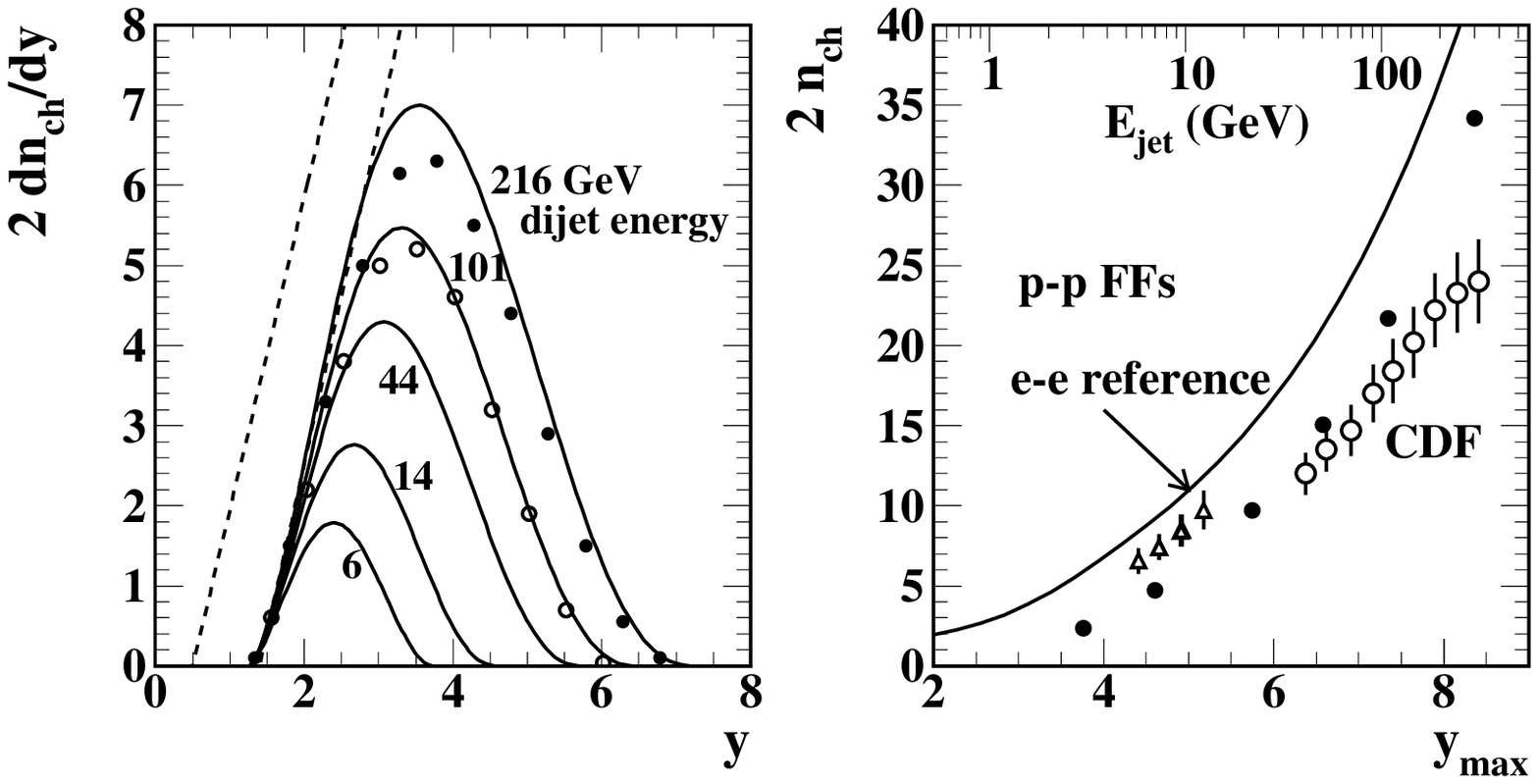}
\caption{\label{eefdb} 
First: Parametrized  $e^+$-$e^-$ FFs for five dijet energies,
Second: Corresponding dijet multiplicities for in-vacuum (solid points) and in-medium (open points) FFs,
Third: Parametrized  p-\=p FFs for five dijet energies compared to CDF data (points)~\cite{cdf2},
Fourth: Corresponding dijet multiplicities for p-p FFs (solid points) and published values (open points~\cite{cdfmult,cdf3}).
} 
\end{figure}

Figure~\ref{eefdb} (first panel) shows parametrized beta FFs for five e-e energy scales. The  $Q = 6$ GeV scale is associated with minijets as explained below. Such curves provide a complete description of e-e FFs at energy scales relevant to nuclear collisions.
Fig.~\ref{eefdb} (second panel) shows light-quark dijet multiplicity systematics from  the same beta parametrization. The solid points correspond to the FFs in the first panel. The open circles represent multiplicities from medium modification of those FFs in central Au-Au collisions at 200 GeV, as described in Sec.~\ref{bweloss}. The ``in-medium'' shift of FFs to smaller fragment momenta requires more fragments to satisfy parton-energy conservation. The systematics of quark and gluon jets coincide for energy scales $Q = 2 E_{jet} < 8$ GeV ($y_{max} < 4$). 

Figure~\ref{eefdb} (third panel) shows e-e beta FFs for five parton energies~\cite{ffprd} modified by the $g_{cut}$ factor to describe p-p FFs.  The deviation from e-e FFs is indicated by the two dotted lines~\cite{evolve}. The CDF data (points) are from~\cite{cdf2}.
Fig.~\ref{eefdb} (fourth panel) shows multiplicity systematics (solid points) for p-p (i.e., modified e-e) FFs from the parametrization. The solid curve represents unmodified e-e FFs as a reference. There is substantial reduction of p-p FF multiplicities due to the cutoff. Also plotted are CDF FF multiplicities from reconstructed jets (open triangles~\cite{cdfmult} and open circles~\cite{cdf3}). 

Comparison of Fig.~\ref{eefdb} second and fourth panels reveals that dijet multiplicities (and charged-particle energy integrals) are strongly suppressed in p-p collisions compared to equivalent FFs in e-e collisions. p-p jet multiplicities are reduced by 30-70\%. FFs are apparently ``modified'' in p-p collisions as well as A-A collisions. At $Q = 6$ GeV (minijets) there is a three-fold dijet multiplicity reduction for p-p relative to e-e collisions.

\section{Parton spectrum model}

A model for the parton $p_t$ spectrum resulting from minimum-bias scattering into an $\eta$ acceptance near projectile mid-rapidity can be parametrized as
\bea
\frac{1}{p_t}\frac{d\sigma_{dijet}}{dp_t} = f_{cut}(p_t) \frac{A_{p_t}}{p_t^{n_{QCD}}} \rightarrow  \frac{d\sigma_{dijet}}{dy_{max}} = f_{cut}(y_{max})\, A_{y_{max}}\, \exp\{-(n_{QCD} - 2)\, y_{max}\},
\eea
which defines QCD exponent $n_{QCD}$, with  $y_{max} \equiv\ln(2\,p_t / m_\pi )$. The cutoff factor
\bea
 f_{cut}(y_{max}) = \{ \tanh[(y_{max} - y_{cut})/\xi_{cut}] + 1\}/2
\eea
represents in this analysis the minimum parton momentum which leads to detectable charged hadrons as neutral pairs (i.e., local charge ordering). Parton spectrum and cutoff parameters are determined via FD comparisons with p-p and Au-Au spectrum hard components.

\begin{figure}[h]
\includegraphics[width=.32\textwidth,height=.27\textwidth]{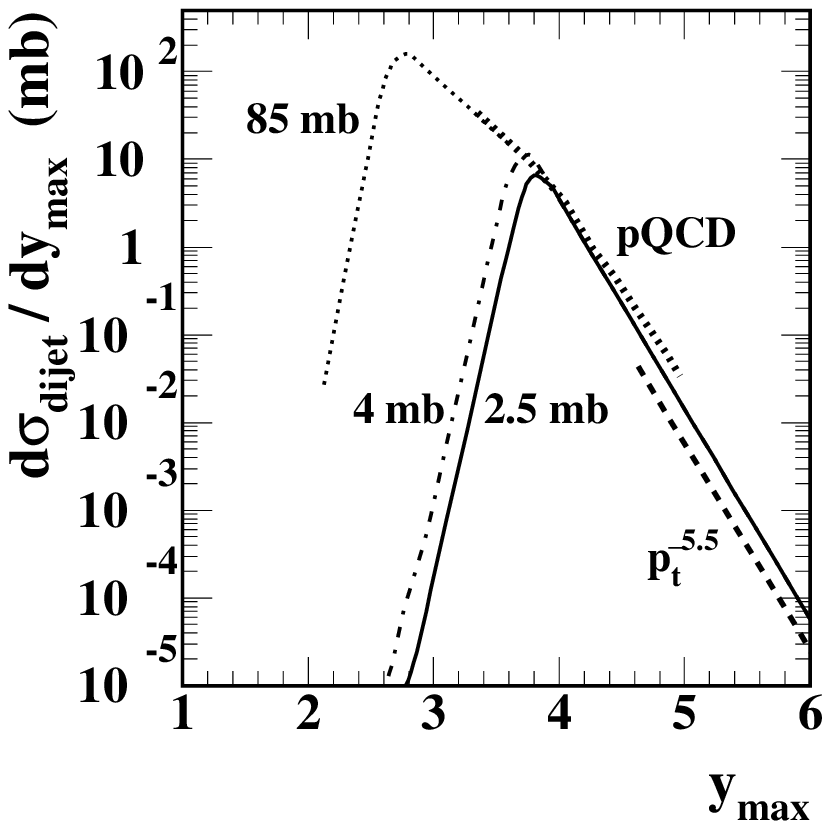}
\includegraphics[width=.32\textwidth,height=.275\textwidth]{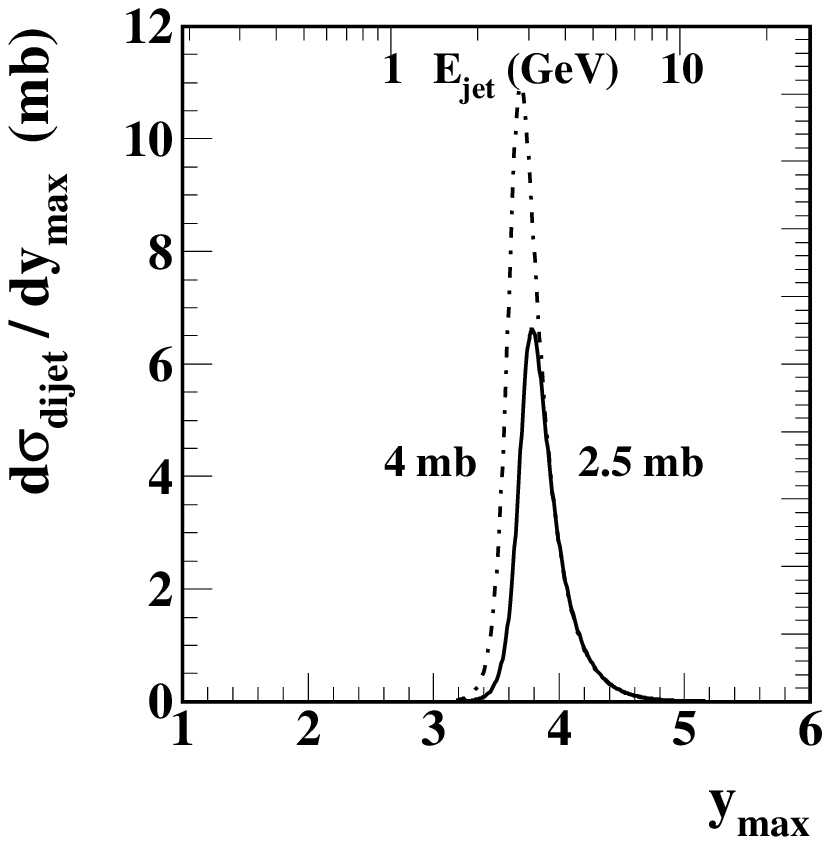}
\includegraphics[width=.32\textwidth,height=.27\textwidth]{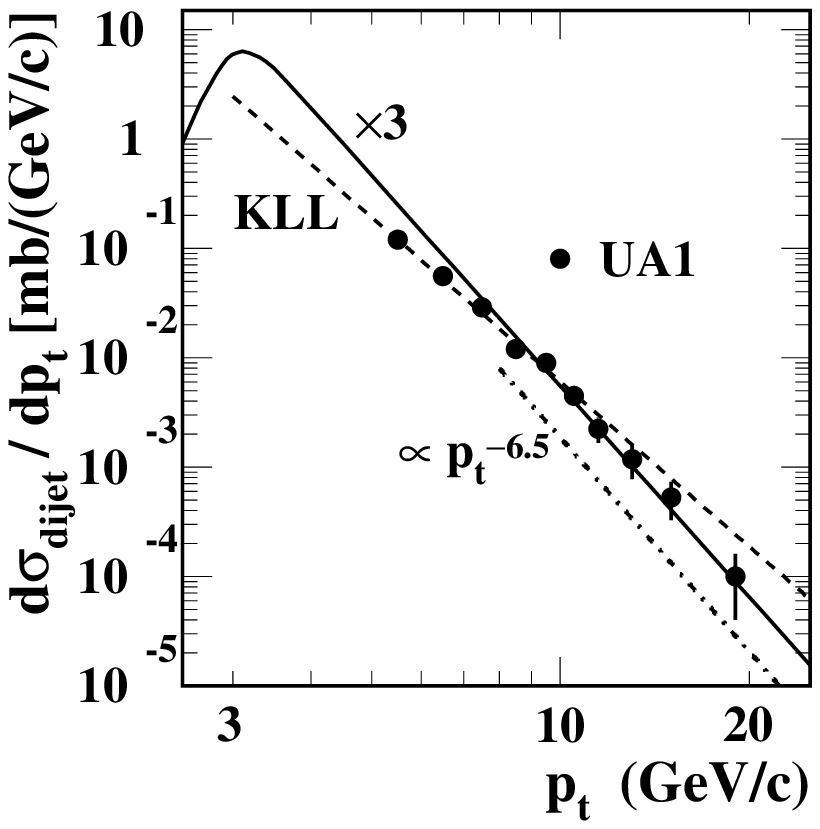}
\caption{\label{partspec} 
First: Parton spectra inferred from this analysis for p-p collisions (solid curve) and central Au-Au collisions (dash-dotted curve) compared to an ab-initio pQCD theory result (bold dotted curve~\cite{cooper}),
Second: Parton spectra from this analysis in a linear plot,
Third: Parton spectrum from reconstructed jets (UA1, solid points~\cite{ua1}) compared to theory (dashed curve~\cite{kll}) and this analysis (solid curve, note factor 3).
} 
\end{figure}

Fig.~\ref{partspec} (semilog and linear formats) shows the parton spectrum (solid curve) inferred from a p-p spectrum hard component~\cite{evolve}. $y_{cut}$ and $A_{y_{max}}$ are well-defined by the p-p hard component, and $n_{QCD}$ is defined by Au-Au spectrum hard components extending to larger $y_t$. The dotted curve in the first panel is an ab-initio pQCD calculation~\cite{cooper}. The linear plot (second panel) indicates the narrowness of the spectrum, with effective mean energy near 3 GeV (minijets).
Fig.~\ref{partspec} (third panel) compares the spectrum defined in this analysis (solid curve, and note the factor 3) with 200 GeV UA1 jet cross-section data obtained by event-wise jet reconstruction~\cite{ua1}. The UA1 spectrum integral is 4 mb~\cite{ua1}.
 The spectrum from this analysis integrates to $2.5 \pm 0.6$ mb with well-defined cutoff $\sim 3$ GeV which agrees well with pQCD theory (e.g.,~\cite{sarc}). The KLL parametrization $600/p_t^5$ mb/(GeV/c) (dashed line) integrates to 2.2 mb above 3 GeV/c~\cite{kll}.

\section{Fragment distributions from a QCD folding integral}

The folding integral used to obtain fragment distributions (FDs) in this analysis is
\bea \label{fold}
\frac{d^2n_{h}}{dy\, d\eta}  \hspace{-.05in} &\approx&  \frac{\epsilon(\delta \eta,\Delta \eta)}{ \sigma_{NSD}\, \Delta \eta}\int_0^\infty \hspace{-.07in}  dy_{max}\, D_{xx}(y,y_{max})\, \frac{d\sigma_{dijet}}{dy_{max}},
\eea
where $D_{xx}(y,y_{max})$ is the dijet FF ensemble  from a source collision system (xx = e-e, p-p, A-A, in-medium or in-vacuum), and $d\sigma_{dijet}/dy_{max}$ is the minimum-bias parton spectrum~\cite{evolve}. Hadron spectrum hard component ${d^2n_{h}}/{dy\, d\eta}$ as defined represents the fragment yield from  scattered parton pairs into one unit of $\eta$. Efficiency factor $\epsilon \sim 0.5$ (for a single dijet and one unit of $\eta$) includes the probability that the second jet also falls within $\eta$ acceptance $\delta \eta$ and accounts for losses from jets near the acceptance boundary. $\Delta \eta \sim 5$ is the effective $4\pi$ $\eta$ interval for scattered partons. $\sigma_{NSD}$ ($\sim 36$ mb for $\sqrt{s_{NN}} = 200$ GeV) is the cross section for NSD p-p collisions.

\begin{figure}[h]
\includegraphics[width=.24\textwidth,height=.244\textwidth]{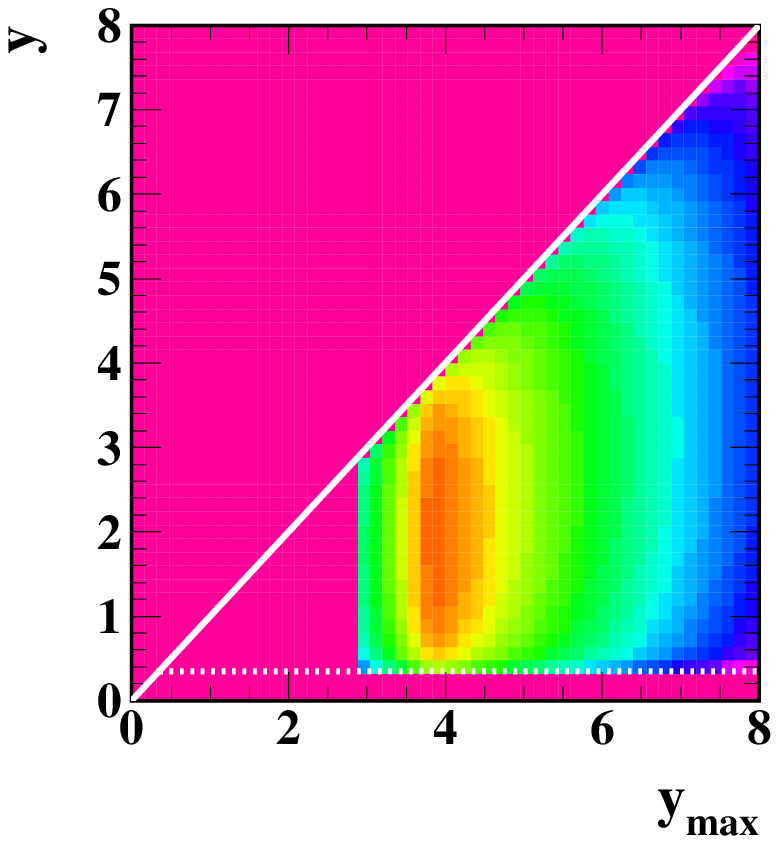}
\includegraphics[width=.24\textwidth,height=.24\textwidth]{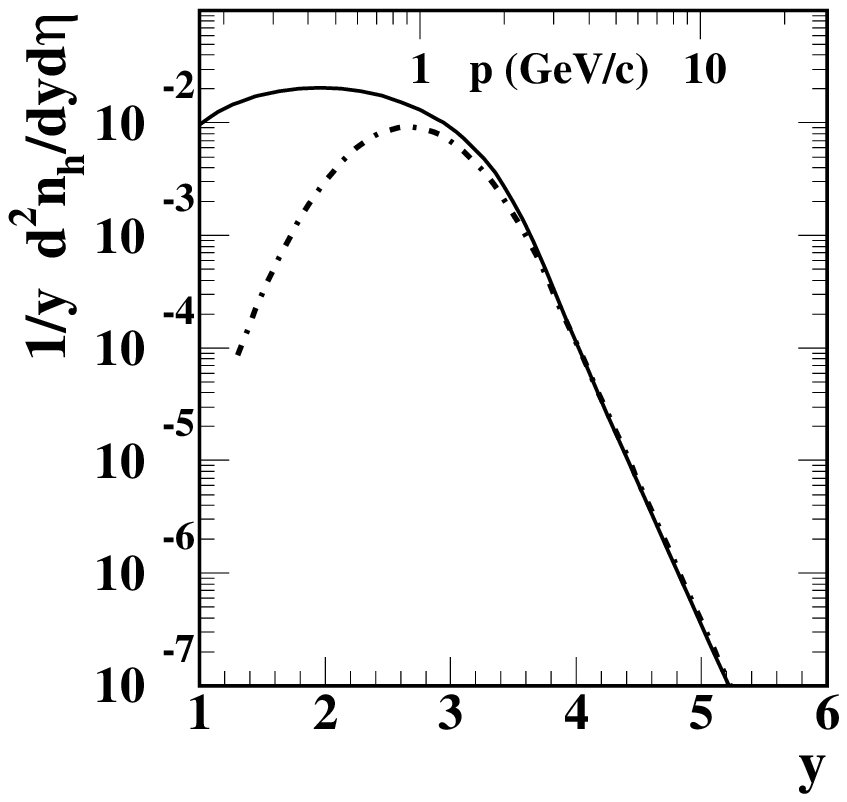}
\includegraphics[width=.24\textwidth,height=.244\textwidth]{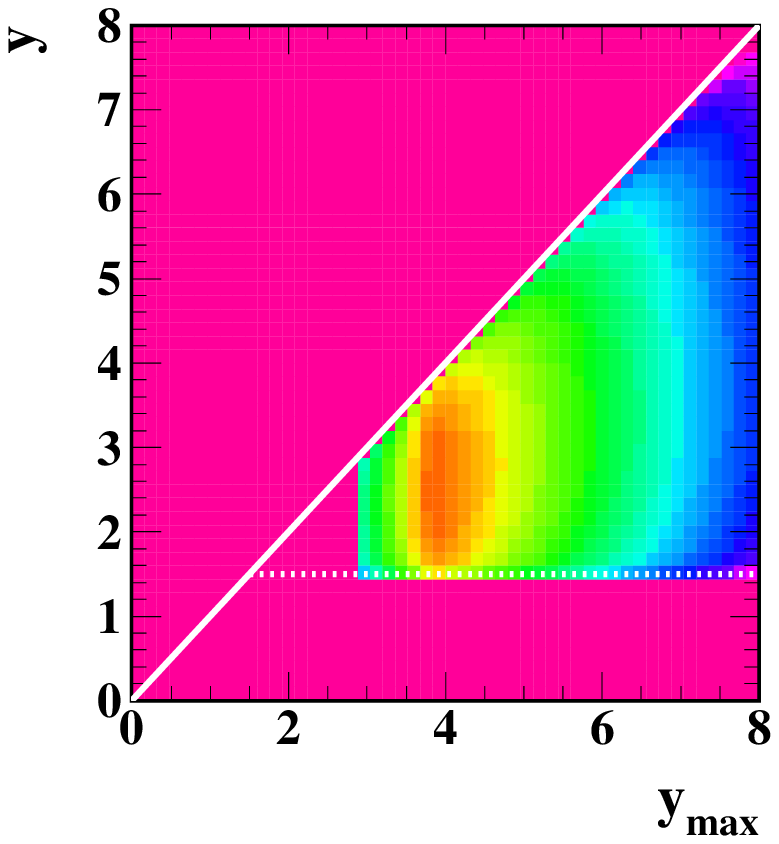}
\includegraphics[width=.24\textwidth,height=.24\textwidth]{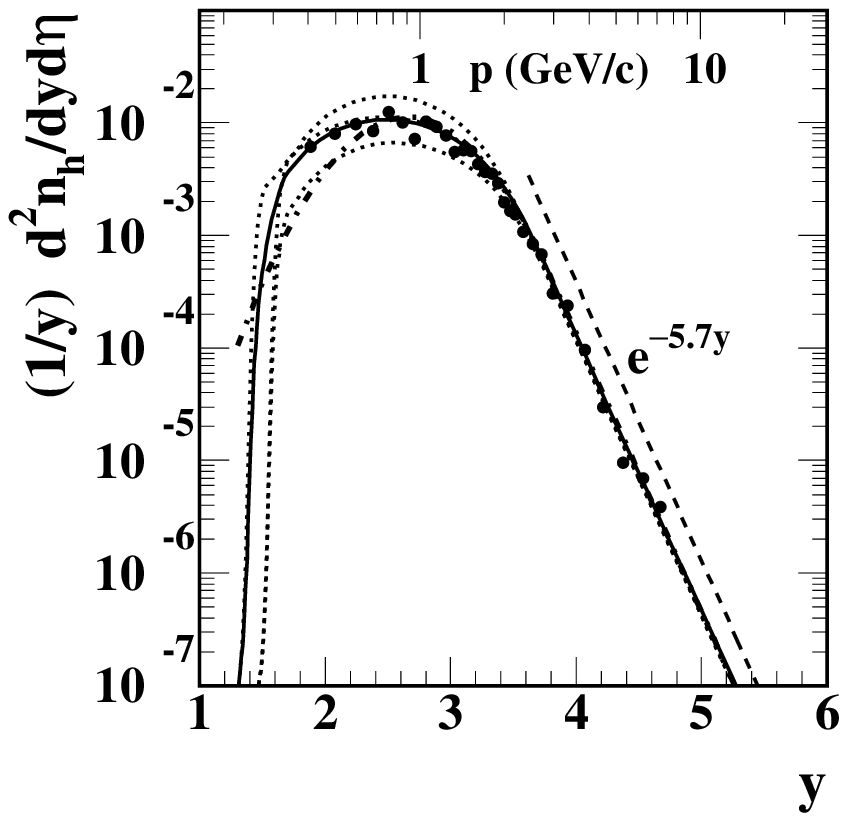}
\caption{\label{fd} 
First: pQCD folding-integral argument for $e^+$-$e^-$ FFs,
Second: Corresponding fragment distribution (solid curve) and p-p hard-component reference (dash-dotted curve),
Third: Folding-integral argument for p-\=p FFs,
Fourth: Corresponding fragment distribution (solid curve) compared to p-p hard-component data (points). Dotted curves correspond to $\pm$10\% change in parton spectrum cutoff energy about 3 GeV. 
} 
\end{figure}

Fig.~\ref{fd} (first panel) shows the integrand $D_{ee}(y,y_{max})\, \frac{d\sigma_{dijet}}{dy_{max}}$ of the folding integral in Eq.~(\ref{fold}) incorporating unmodified FFs from e-e collisions with lower bound at  $y_{min} \sim 0.35$ ($p_t \sim 0.05$ GeV/c) (dotted line). The plot $z$ axis is logarithmic to show structure over the entire distribution support.
Fig.~\ref{fd} (second panel) shows  the corresponding FD (solid curve). The parton spectrum parameters determined from the p-p hard component are retained. The solid curve is the ``correct answer'' for an FD describing inclusive hadrons from inclusive partons produced by {\em free} parton scattering from p-p collisions, which is not observed in real nuclear collisions. The dash-dotted curve represents the hard-component model inferred from p-p collisions~\cite{ppprd}. The FD from e-e FFs lies well above the measured p-p hard component for hadron $p_t < 2$ GeV/c ($y_t < 3.3$), and the mode is shifted down to $\sim 0.5$ GeV/c. The ``correct'' e-e FD strongly disagrees with the relevant part of the p-p $p_t$ spectrum---the hard component. Despite the strong disagreement the e-e FD is the correct reference for nuclear collisions, as demonstrated below.

Fig.~\ref{fd} (third panel) shows a surface plot of integrand $D_{pp}(y,y_{max})\, \frac{d\sigma_{dijet}}{dy_{max}}$, incorporating e-e FFs based on the LEP parametrization but modified by the FF cutoff function inferred from p-\=p collisions. The main difference from e-e FFs is that the lower bound of p-p FFs is raised to $y_{min} \sim 1.5$ ($p_t \sim 0.3$ GeV/c from 0.05 GeV/c).
Fig.~\ref{fd} (fourth panel) shows the corresponding FD $H_{NN-vac}$ (integration of the third panel over $y_{max}$) as the solid curve. The mode of the FD is $\sim 1$ GeV/c. The dash-dotted curve is a Gaussian-plus-tail model function, and the solid points are hard-component data from p-p collisions~\cite{ppprd}. That comparison determines parton spectrum parameters $y_{cut} = 3.75$ ($E_{cut} \sim 3$ GeV), $A_{y_{max}}$ and exponent $n_{QCD} = 7.5$. The p-p data are well-described by the pQCD folding integral.
This procedure establishes an absolute quantitative relationship among parametrized parton spectrum, measured FFs and measured spectrum hard components over all $p_t$, not just a restricted interval (e.g., above 2 GeV/c).

\section{Parton ``energy loss'' and medium-modified FDs} \label{bweloss}

The hypothesis of parton energy loss in a thermalized bulk medium is of central importance at RHIC. In some models the medium is opaque to most hard-scattered partons -- only a small fraction emerge as correlated fragments. But minijet systematics suggest no parton loss to thermalization. In this section I adopt a pQCD-inspired minimal model of FF modification (Borghini-Wiedemann or BW)~\cite{bw}, with no loss of parton energy to a medium or scattered partons to thermalization.

\begin{figure}[h]
\includegraphics[width=.24\textwidth,height=.244\textwidth]{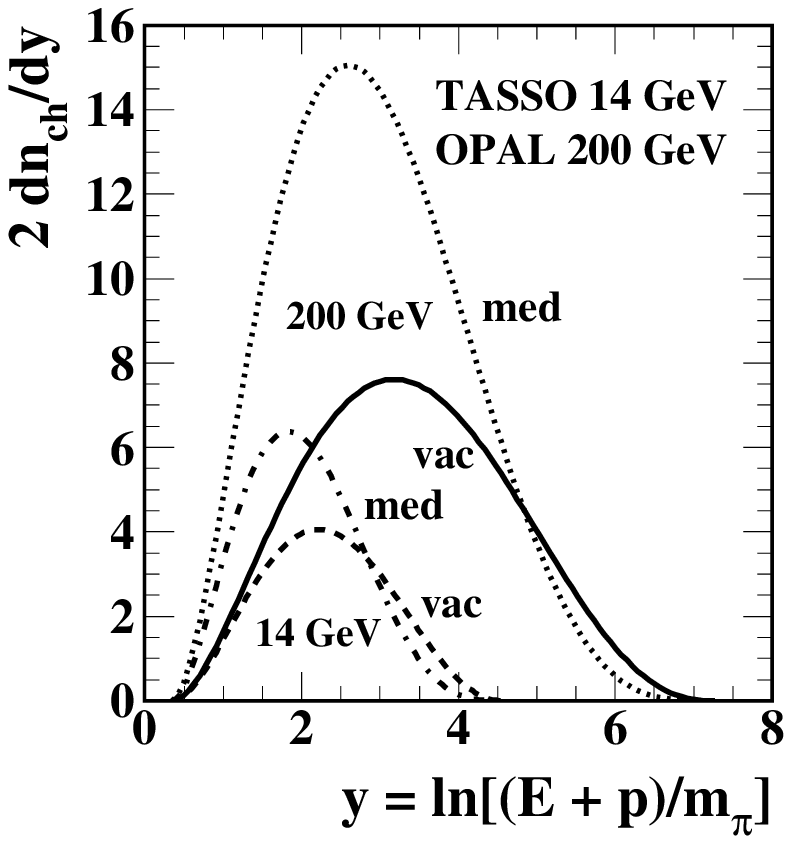}
\includegraphics[width=.24\textwidth,height=.24\textwidth]{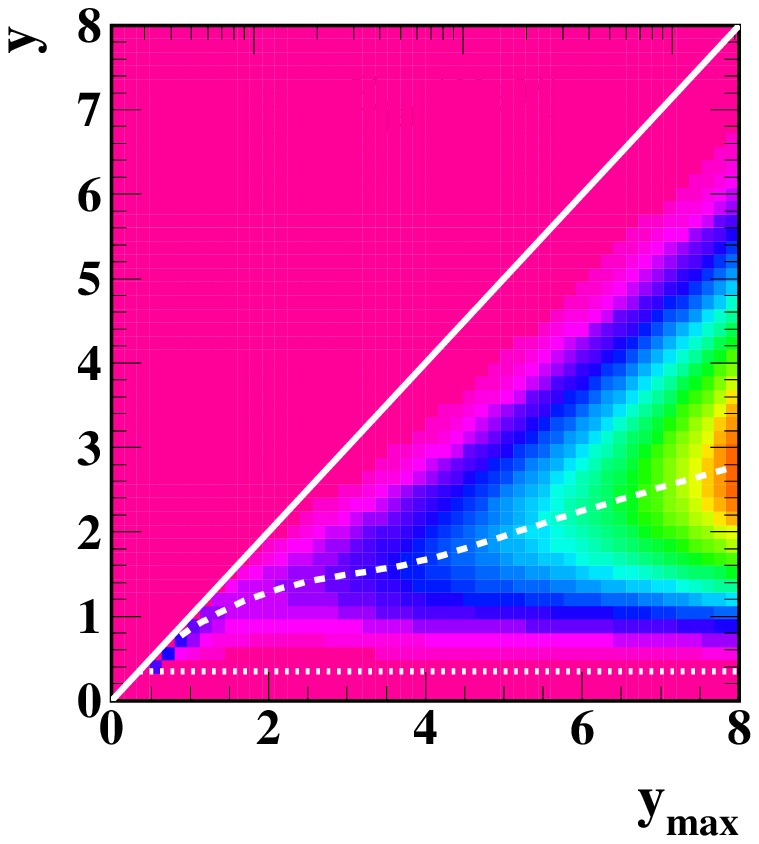}
\includegraphics[width=.24\textwidth,height=.244\textwidth]{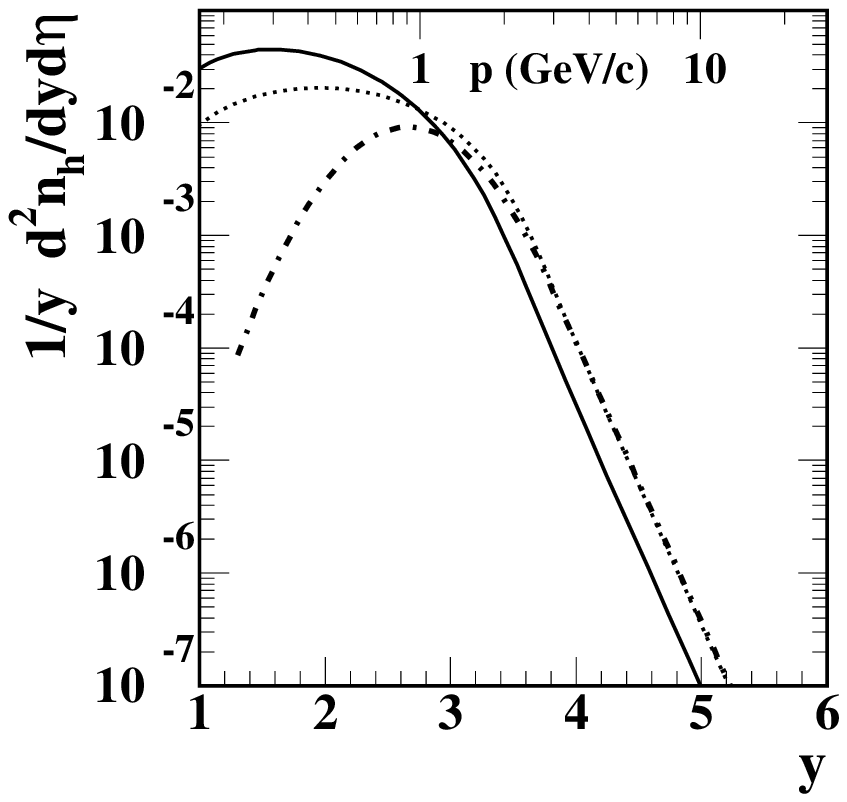}
\includegraphics[width=.24\textwidth,height=.24\textwidth]{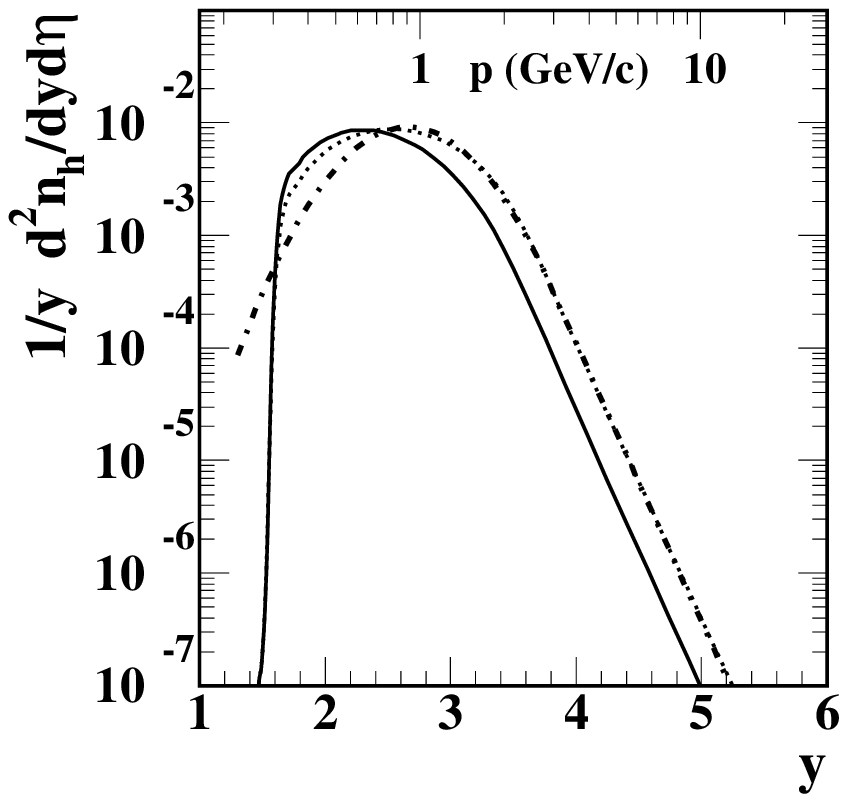}
\caption{\label{eloss1} 
First: $e^+$-$e^-$ FFs for two energies unmodified (solid and dashed curves) and modified according to a rescaling procedure~\cite{bw} (dash-dotted and dotted curves) to emulate parton ``energy loss,''
Second: $e^+$-$e^-$ FF ensemble modified according to~\cite{bw},
Third: Medium-modified FD from $e^+$-$e^-$ FFs (solid curve) compared to in-vacuum $e^+$-$e^-$ FD (dotted curve)
Fourth: Medium-modified FD from p-\=p FFs (solid curve) compared to in-vacuum FD (dotted curve).
} 
\end{figure}

 
Figure~\ref{eloss1} (first panel) illustrates the BW model of FF modification (cf. Fig. 1 of~\cite{bw}). In-vacuum e-e FFs for $Q = 14$ and 200 GeV from the beta parametrization are shown as dashed and solid curves respectively~\cite{ffprd}. Whereas the BW model was expressed on $\xi_p$ FFs are plotted here on fragment rapidity $y$. The relation is  $\xi_p = \ln(p_{jet}/ p)$ = $\ln(2\, p_{jet} / m_\pi) - \ln(2p/m_\pi) \sim y_{max} - y$, with energy scale $Q = 2\, p_{jet}$. The practical consequence of the BW ``energy-loss'' mechanism is a momentum-conserving rescaling of FFs on $x_p$, with $\xi_p = \ln(1/x_p)$. Small density reductions at larger fragment momenta (smaller $\xi_p$) are compensated by much larger increases at smaller momenta. The largest changes (central Au-Au) correspond to an inferred  25\% leading-parton fractional ``energy loss.'' We model the BW modification simply by changing parameter $q$ in $\beta(u;p,q)$ by $\Delta q \sim 1$, which accurately reproduces the BW result. The modified FFs are the dash-dotted and dotted curves~\cite{evolve}. Fig.~\ref{eloss1} (second panel) shows the modified e-e FF ensemble with FF modes shifted to smaller fragment rapidities. {\em No energy is lost from FFs in this model}.

Figure~\ref{eloss1} (third panel) shows $H_{ee-med}$ (solid curve), the FD obtained by inserting e-e in-medium FFs from the second panel into Eq.~(\ref{fold}) and integrating over parton rapidity $y_{max}$. The dotted curve is the $H_{ee-vac}$ reference from in-vacuum e-e FFs. The dash-dotted curve is again the Gaussian-plus-tail p-p hard component $H_{GG}$ reference. The mode of $H_{ee-med}$ is $\sim 0.3$ GeV/c.
Fig.~\ref{eloss1} (fourth panel) shows results for p-p FFs. Major differences between p-p and e-e FDs appear below $p_t \sim 2$ GeV/c ($y_t \sim 3.3$). Conventional comparisons with theory (e.g., data {\em vs} NLO FDs) typically do not extend below 2 GeV/c~\cite{phenixnlo}. The large difference between the two systems below 2 GeV/c reveals that the small-$p_t$ region, conventionally assigned to hydro phenomena, may be of central importance for understanding fragmentation evolution in A-A collisions.

\section{Fragment evolution with centrality in Au-Au collisions}

\begin{figure}[h]
\includegraphics[width=.24\textwidth,height=.244\textwidth]{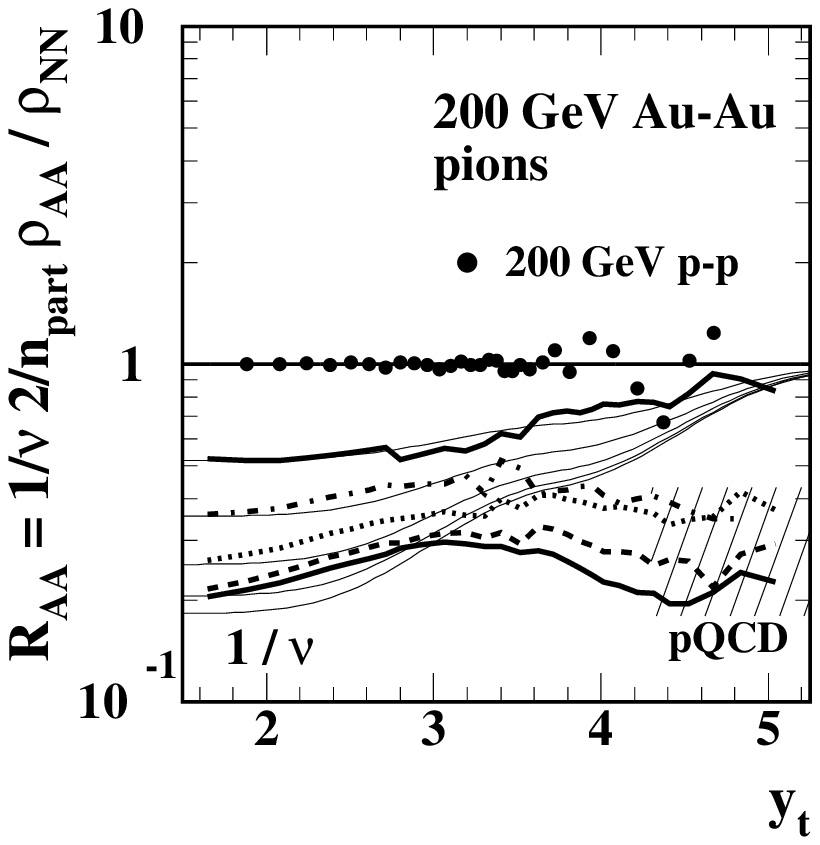}
\includegraphics[width=.24\textwidth,height=.24\textwidth]{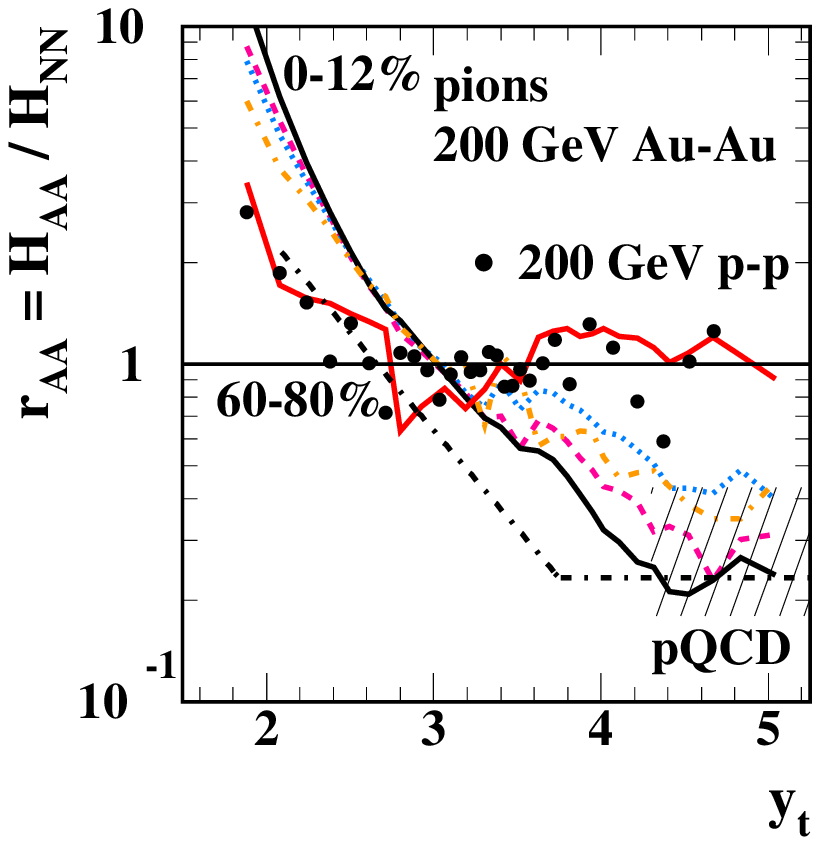}
\includegraphics[width=.24\textwidth,height=.244\textwidth]{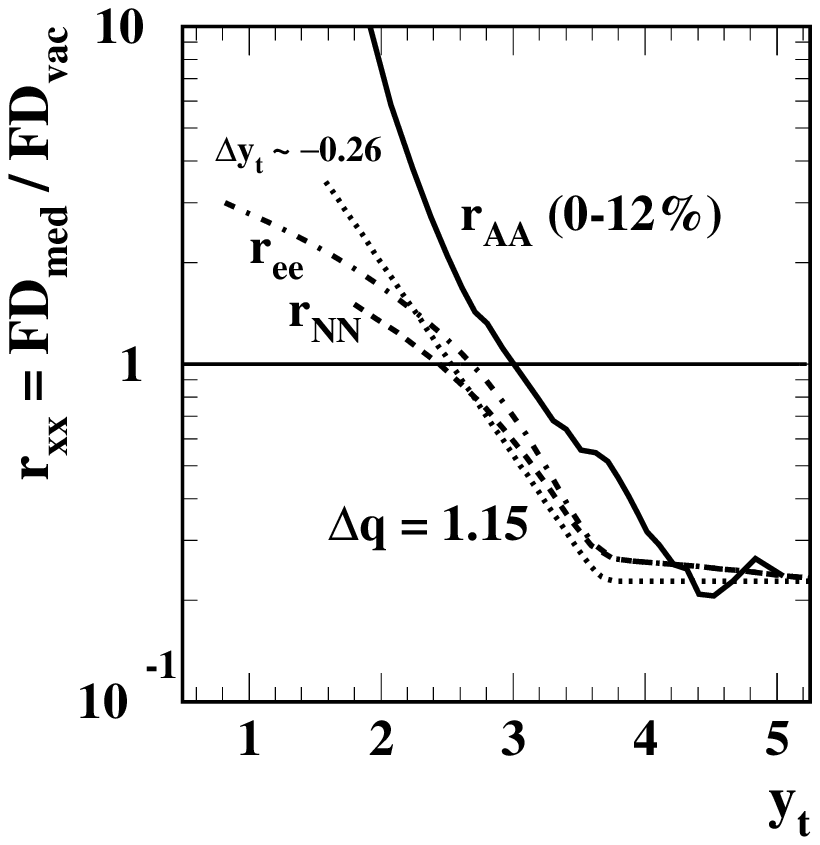}
\includegraphics[width=.24\textwidth,height=.24\textwidth]{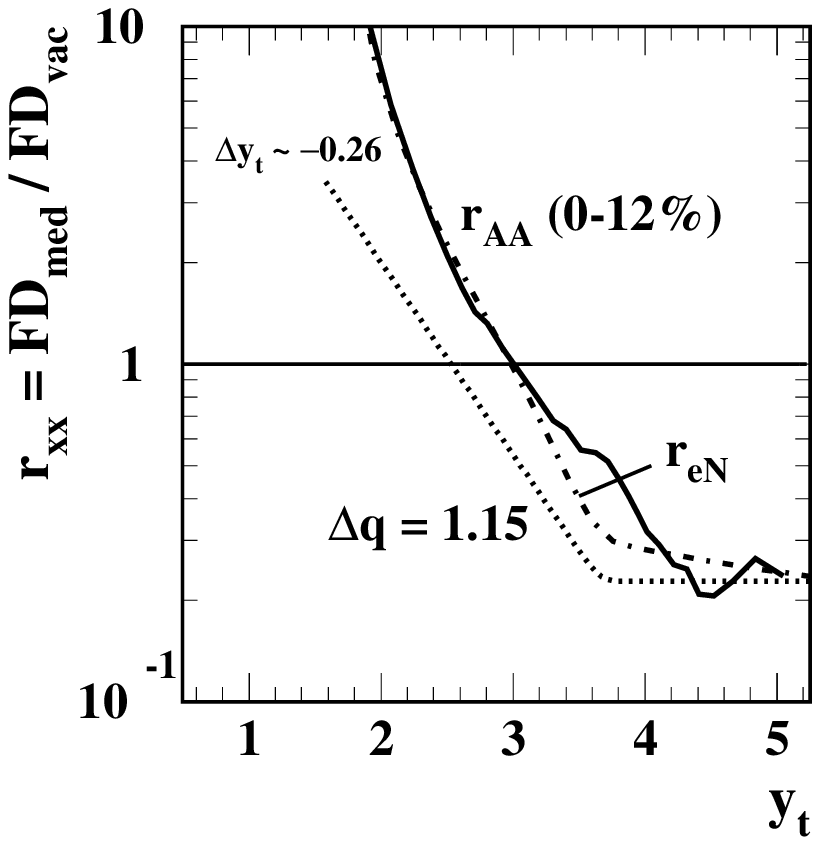}
\caption{\label{evo1} 
First: Conventional spectrum-ratio measure $R_{AA}$, illustrating strong suppression of spectrum information below 4 GeV/c ($y_t = 4$),
Second: Hard-component ratio $r_{AA}$ illustrating restoration of suppressed structure at small $y_t$,
Third: Comparison of calculated FD ratios to measured $r_{AA}$ for central Au-Au collisions,
Fourth: Comparison of novel FD ratio $r_{eN}$ to measured $r_{AA}$ for central Au-Au collisions.
} 
\end{figure}

We have established a system to combine measured FFs and a parametrized pQCD parton spectrum to produce calculated fragment distributions $FD_{xx}$ for comparison with measured spectrum hard components $H_{xx}$. Conventional comparisons employ a ratio measure. Two questions emerge: what is the validity of the ratio definition, and what should be the reference for such a ratio.
%
%
The conventional spectrum ratio at RHIC is $R_{AA}$, defined in the first line of
\bea \label{raa}
R_{AA} &\equiv& \frac{1}{\nu} \times \frac{S_{NN}(y_t) + \nu\, H_{AA}(y_t;\nu)}{S_{NN}(y_t) +  H_{NN}(y_t)} \\ \nonumber
&\rightarrow& \frac{1}{\nu} + \frac{H_{NN}}{S_{NN}}\, r_{AA} ~~~~~~{\rm at}~~~ y_t = 2.
 \eea
In that definition the terms in numerator and denominator are normalized per participant pair $n_{part} / 2$, so the prefactor is $1/\nu$ rather than $1/n_{binary}$.
Fig.~\ref{evo1} (first panel) illustrates problems with that measure. Hard-component evolution with centrality, the main object of this analysis, is described by ratio $r_{AA} \equiv H_{AA} / H_{NN}$.  The second line of Eq.~(\ref{raa}) gives the limiting value of $R_{AA}$ near $y_t \sim 2$ where the ${H_{NN}}/{S_{NN}}$ ratio is typically $\sim 1/170$. $r_{AA}$ is thus suppressed by a large factor in just the interval where fragmentation details are most important. The p-p data (dots) illustrate suppression of even statistical fluctuations. All information is lost.

Figure~\ref{evo1} (second panel) shows ratio $r_{AA}$ based on hard-component reference $H_{NN}$ set equal to Gaussian model $H_{GG} = n_h\, H_0$ from~\cite{ppprd}. Evolution of suppression and enhancement is dramatically more accessible. The p-p data and the most peripheral Au-Au data agree with the N-N reference ($r_{AA} = 1$) above $y_t = 2.5$ but deviate significantly from $H_{GG}$ below that point.
For the Au-Au collisions in this figure $\nu \equiv 2\, n_{bin} / n_{part}$ values for five centralities are 1.93, 2.83, 3.92, 4.87, 5.5, where $\nu \sim 1.25$ is N-N collisions and $\nu \sim 6$ is $b = 0$ Au-Au collisions~\cite{hardspec}. From $\nu$ = 1.98 to $\nu$ = 2.83 there is a dramatic change in the hard component. At the transition point $\nu \sim 2.5$ $n_{part} = 40$ (out of 382) and $n_{bin} = 50$ (out of 1136).

Figure~\ref{evo1} (third panel) shows calculated FD ratios $r_{xx} = FD_{xx-med}/FD_{xx-vac}$ with $xx =$ e-e (dash-dotted curve, e-e FFs) or N-N (dashed curve, p-p FFs)~\cite{evolve}.  The solid curve is the measured $r_{AA}$ from central (0-12\%) Au-Au collisions at 200 GeV~\cite{hardspec}. $\Delta q \sim 1.15$ for $H_{ee-med}$ and $H_{NN-med}$ (in-medium FFs) was adjusted to obtain the correct large-$y_t$ suppression for 0-12\% central Au-Au. The reference for $r_{AA}$ is hard-component model function $H_{GG}$. The dotted curve is a reference ratio obtained by shifting $H_{GG}$ on $y_t$ by $\Delta y_t \sim -0.26$ (negative boost)~\cite{hardspec}. The simple negative-boost model does not describe the Au-Au data. But the e-e and N-N ratios also do not describe the data.

Figure~\ref{evo1} (fourth panel) introduces a novel concept. Instead of comparing the calculated in-medium FD for N-N collisions averaged within A-A collisions with the in-vacuum FD for isolated N-N collisions, or similarly comparing e-e with e-e as in the third panel, the in-medium FD for e-e is compared with the  in-vacuum FD for N-N by defining ratio
\bea
r_{eN} &=& \frac{FD_{ee-med}}{FD_{NN-vac}}.
\eea
Calculated $r_{eN}$ describes the measured $r_{AA}$ well over the entire fragment momentum range. We conclude that $FD_{NN-vac}$ is not the correct reference. The proper  in-vacuum reference for all systems is an FD from e-e FFs, not p-p FFs. We define FD ratios $r_{xx} = FD_{xx-yyy}/FD_{ee-vac}$ with xx = ee, NN, AA and yyy = med or vac to be compared with equivalent spectrum hard components $H_{xx-yyy}$.

\begin{figure}[h]
\includegraphics[width=.65\textwidth,height=.3\textwidth]{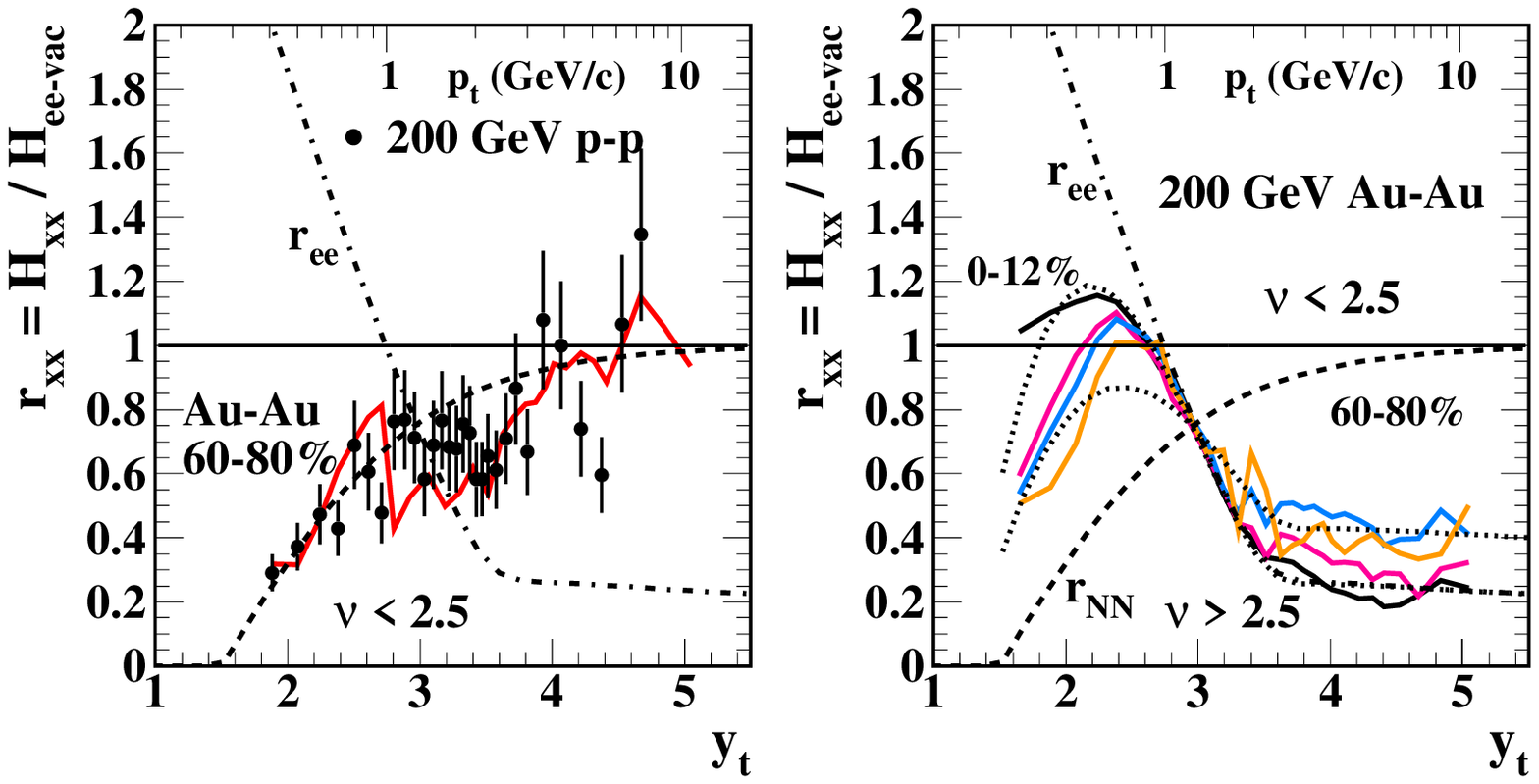}
\includegraphics[width=.33\textwidth,height=.295\textwidth]{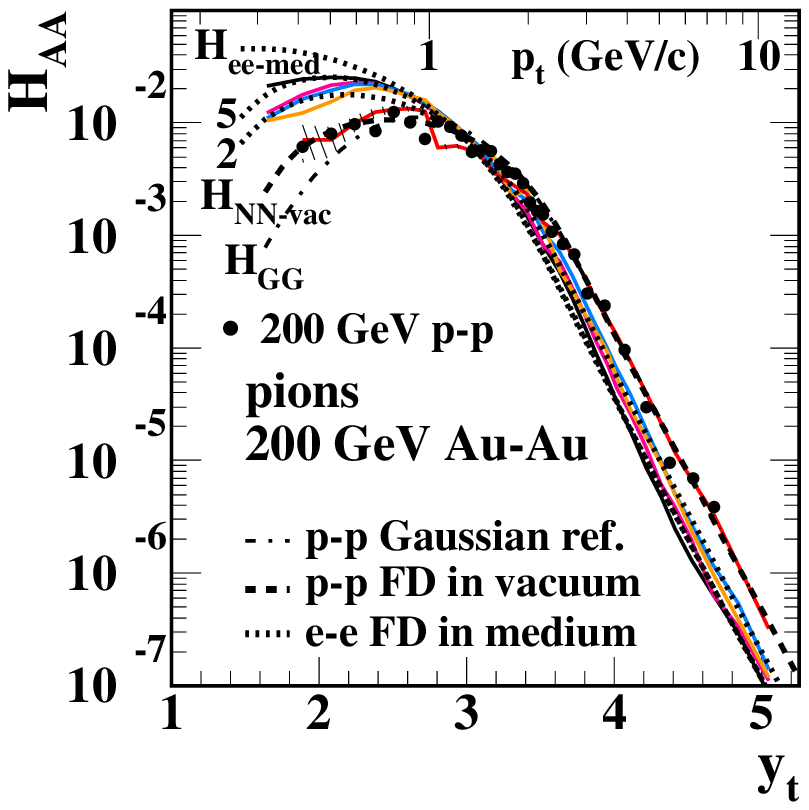}
\caption{\label{evo2} 
First: FD ratios relative to an ee-vacuum reference for Au-Au collisions below the sharp transition,
Second: FD ratios relative to an ee-vacuum reference for Au-Au collisions above the sharp transition revealing major changes in FD structure,
Third: Hard-component evolution in central Au-Au collisions vs centrality. Large increases in fragment yield at smaller $y_t$ ($p_t < 2$ GeV/c) accompany suppression at large $y_t$.
} 
\end{figure}

Figure~\ref{evo2} (first panel) shows ratios redefined in terms of the ee-vac reference: $H_{pp}$ (p-p data -- points),  $H_{AA}$ (peripheral Au-Au data -- solid curve) and calculated $H_\text{ee-med}$ (dash-dotted curve) and $H_{NN-vac}$ (dashed curve) all divided by reference $H_{ee-vac}$. The strong suppression of p-p and peripheral Au-Au data apparent at smaller $y_t$ results from the cutoff of p-p FFs noted above. The comparison is linear rather than logarithmic, as in Fig.~\ref{evo1}, and is thus more differential.

Figure~\ref{evo2} (second panel) shows measured $H_\text{AA}/H_\text{ee-vac}$ for more-central Au-Au collisions (solid curves) above a transition point on centrality at $\nu \sim 2.5$. The main difference is partial restoration of the suppressed region at smaller $y_t$ and suppression at larger $y_t$. The latter has been the major observation at RHIC for jet-related modification (high-$p_t$ suppression, ``jet quenching''~\cite{starraa}). Apparent from this analysis is the accompanying very large {\em increase} in fragment yield {\em below} 2 GeV/c, still strongly correlated with the parent parton~\cite{daugherity}. Also notable is the substantial gap between the peripheral data and the four more-central spectra~\cite{hardspec}. Changes in fragmentation depend very strongly on centrality near the transition point. It is remarkable that the trend at 10 GeV/c corresponds closely to the trend at 0.5 GeV/c. Calculated FD ratio $r_{ee}$ (dash-dotted curve) corresponds to a parton spectrum cutoff shifted down to 2.7 GeV from 3 GeV for p-p collisions, as shown in Fig.~\ref{partspec} (first and second panels). The shift may result from an increased hadron density of states~\cite{evolve}.

Figure~\ref{evo2} (third panel) shows spectrum hard components $H_{AA}$ (solid curves) for five centralities from 200 GeV Au-Au collisions~\cite{hardspec}. The hard components of $y_t$ spectra scale proportional to $n_{binary}$, as expected for parton scattering and fragmentation in A-A collisions (jets). The points are hard-component data from 200 GeV NSD p-p collisions~\cite{ppprd}. The dash-dotted curve is the standard Gaussian+tail model function $H_\text{GG}$. Calculated FDs are also shown. The dashed curve is $H_\text{NN-vac}$, and the upper dotted curve is $H_\text{ee-med}$ with $\Delta q = 1.15$, which nominally corresponds to the most-central Au-Au curve (0-12\%). The parton spectrum cutoff for $H_\text{ee-med}$ has been reduced from 3 GeV ($y_{max} = 3.75$) to 2.7 GeV ($y_{max} = 3.65$) to match the central Au-Au hard component near $y_t = 3$. The dotted curves labeled 2 and 5 (Au-Au centralities) are $H_\text{ee-med}$ with cutoff parameters $y_0 = \xi_y$ reduced to accommodate the data below $y_t = 2.5$. $H_{pp}$, $H_{AA}$ and ratios based on the e-e in-vacuum reference are thus well described by pQCD FD ratio data from 0.3 to 10 GeV/c~\cite{evolve}.

\section{Discussion}

This analysis establishes a quantitative correspondence between calculated pQCD FDs and measured spectrum hard components $H_{xx}$ over the entire fragment $p_t$ range and parton spectrum. We obtain direct access to medium-modified  FFs and the underlying parton spectrum. 

In p-p and in peripheral Au-Au collisions below a transition point at $\nu \sim 2.5$ the underlying power-law parton spectrum terminates near 3 GeV. Hard component $H_{pp}$ or $H_{AA}$ is strongly suppressed at smaller $y_t$ (jet bases excluded from the acceptance) corresponding to p-\=p FFs. The suppression mechanism may be hard-Pomeron (color singlet) exchange in N-N collisions leading to color connections different in p-p than in e-e collisions (which produce q-\=q color dipoles).

Above the transition point: 1) Measured $H_{AA}$ is strongly enhanced at smaller $y_t$ (FF bases partially restored) but suppressed at larger $y_t$ (so-called ``jet quenching''), as observed in~\cite{hardspec}. 2) Corresponding calculated FDs can be generated by incorporating a ``medium-modified'' e-e FF scenario---simple rescaling of e-e splitting functions---which implies a three-fold increase in jet multiplicity compared to p-\=p FFs. 3) The parton spectrum cutoff is reduced, by up to 10\% in central Au-Au collisions implying a 50\% increase in the jet cross section and minijet production. 

Evolution of $H_{AA}$ corresponds to two-particle correlations on $(y_t,y_t)$~\cite{daugherity}. Observed spectrum hard-component systematics indicate that no partons are ``absorbed'' or lost to thermalization (no ``opaque core'' is formed). All scattered partons predicted by a pQCD differential cross section produce jet-correlated hadrons in the final state. The minimum-bias jet fragment yield in central Au-Au collisions fully accounts for the increase of collision multiplicity beyond participant scaling (soft component). There is also no indication from correlations, spectrum structure or integrated $p_t$ that parton spectra extend down to 1 GeV as suggested by saturation-scale arguments~\cite{cooper,eskola}.

\section{Summary}

Two-component decomposition of hadron spectra from p-p and Au-Au collisions isolates minimum-bias parton fragment distributions as spectrum hard components ($H_{xx}$) which can be estimated theoretically by folding measured fragmentation functions (FFs) with a pQCD parton spectrum to produce calculated fragment distributions (FDs). In this analysis accurate parameterizations of p-\=p and $e^+$-$e^-$ FFs for a large range of parton energies are folded with a power-law parton spectrum with cutoff to produce calculated FDs which are compared with measured spectrum hard components from p-p collisions and from Au-Au collisions for several centralities.

Comparisons reveal that FFs in p-p collisions are strongly suppressed for smaller fragment momenta (jet base suppressed). The suppression is possibly related to hard-Pomeron exchange and resulting color-field deviations from q-\=q. Comparisons further indicate that above a specific Au-Au centrality (transition point) there is evolution toward e-e FFs as an asymptotic limit (jet base partially restored). FFs are modified consistent with alteration of parton splitting. No partons are lost to absorption or thermalization (no ``opaque core''), and no significant parton energy is lost from integrated FFs. Perturbative QCD describes parton scattering and fragmentation in nuclear collisions over a large kinematic domain, and minijets dominate collision dynamics in all cases. The most dramatic alteration of parton fragmentation in A-A collisions occurs below $p_t = 2$ GeV/c.

This work was supported in part by the Office of Science of the U.S. DoE under grant DE-FG03-97ER41020

\end{document}